\documentclass[a4paper,10pt]{article}
\pdfoutput=1

\usepackage{jcappub}

\usepackage{graphicx,amsmath,float,natbib,pstricks}

\usepackage[caption=false]{subfig}

\begin{document}

\title{On the visible size and geometry of aggressively expanding civilizations at cosmological distances}

\author{S. Jay Olson}
\affiliation{Department of Physics, Boise State University, Boise, Idaho 83725, USA}
\emailAdd{stephanolson@boisestate.edu}

\abstract{If a subset of advanced civilizations in the universe choose to rapidly expand into unoccupied space, these civilizations would have the opportunity to grow to a cosmological scale over the course of billions of years.  If such life also makes observable changes to the galaxies they inhabit, then it is possible that vast domains of life-saturated galaxies could be visible from the Earth.  Here, we describe the shape and angular size of these domains as viewed from the Earth, and calculate median visible sizes for a variety of scenarios.  We also calculate the total fraction of the sky that should be covered by at least one domain. In each of the 27 scenarios we examine, the median angular size of the nearest domain is within an order of magnitude of a percent of the whole celestial sphere.  Observing such a domain would likely require an analysis of galaxies on the order of a Gly from the Earth.}

\maketitle

\section{Introduction}

At first glance, the universe beyond the Earth seems to be devoid of the influence of advanced life.  This could be because our corner of the universe literally is devoid of advanced life, or it could be that the nature of advanced life is such that one has to look very hard to see any evidence of its existence, even if it is widespread and nearby.  Both possibilities have long been discussed~\cite{hart1975, newman1981}, but the overwhelming majority of SETI searches have corresponded to an exploration of the second possibility~\cite{shuch2011}, that of common advanced life with very low impact on nature. 

The opposite possibility, that advanced life is very rare but very powerful, also deserves consideration.  Indeed, it is important to consider this possibility because there exists a small set of plausible technologies that imply a single planet could come to saturate truly vast domains of the universe with intelligent life, making changes to galaxies that could be visible from cosmological distances.  A sufficient set of technologies includes high speed intergalactic space travel~\cite{armstrong2013}, self-replicating spacecraft~\cite{freitas1980}, artificial intelligence~\cite{bracewell1960}, and Dyson swarms~\cite{dyson1960}.  If such technology is available, the only remaining requirement is a mild preference for making use of the universe's resources, as the cost of initiating such a cosmic engineering effort is tiny -- the cost is exactly that of producing the first self-replicating spacecraft.  It has been argued that expansionistic behavior of this kind can be expected of rational agents in pursuit of an extremely wide range of goals, even if the direct benefit of continued expansion is subject to steeply diminishing returns~\cite{bostrom2014}.

It was quickly recognized in the literature (by both sides of a famous SETI dispute~\cite{tipler1980, sagan1983}) that self-replicating spacecraft could come to fully explore and utilize the resources of the Milky Way galaxy on a very short timescale~\cite{valdes1980}, but the implications at an intergalactic scale have only recently been considered.  In the context of homogeneous cosmology, we have introduced a model to describe the fraction of galaxies in the universe that should have been reached, fully colonized, and engineered to the maximum practical extent (we call this condition ``life-saturated" or just ``saturated"), given the appearance rate of aggressively expanding civilizations in the universe and their expansion speed~\cite{olson2014}.  Such models do not assume that every advanced civilization will choose to expand so rapidly -- only that some non-zero fraction of advanced life will do so.  Further results imply that the distance to the nearest of these expanding, life-saturated domains is likely to be cosmological (at least as distant as the homogeneity scale of the universe) -- a result that follows from the assumption that humanity has arrived at a typical time relative to other, comparable civilizations~\cite{olson2015a}.

Here, we explore an observational signature of these civilizations by calculating the shape and solid angle size of their rapidly expanding domains, as viewed from the Earth -- if they are out there, we want to know the kind of geometry to look for.  We will consider expansion speeds from $.1 c$ to $.9c$, and scenarios in which humanity has arrived at the mean cosmic time for comparable life, or as a two standard deviation latecomer (specifying the relative appearance time of humanity is a way of fixing the rate of appearance of the expanding civilizations).  We also generate a set of scenarios by using a maximum likelihood estimate to set the appearance rate.  In every case where we expect to see any expanding domains at all, we find that the median solid angle size of the nearest visible domain is within an order of magnitude of $1 \%$ of the entire celestial sphere.  Although these domains would most likely be at cosmological distances, their great visible size suggests search techniques that would be far easier to perform than deep, full-sky surveys.

This paper will not investigate the specific observable changes to the individual galaxies within an expanding, life-saturated domain.  This question is much more sensitive to the details of the technology employed by advanced civilizations than the geometric questions we examine here (where technology effects geometry only through the civilization's expansion velocity).  For example, extensive use of Dyson swarms in a life-saturated galaxy would seem to decrease visible wavelength emissions, while increasing infrared emissions (waste heat), with the total galaxy luminosity remaining constant.  However, if it is possible to construct heat engines powered by the Hawking radiation of microscopic, artificially-created black holes (bypassing the limitations inherent in using nuclear fusion as an energy source), Dyson swarms might be a mostly irrelevant technology.  The emissions of life-saturated galaxies in this case might be unchanged in the visible spectrum, while substantially brighter in the part of the spectrum corresponding to waste radiation.  One might also consider the consequences of stellar deconstruction, dust harvesting, or engineering galactic orbits on a massive scale -- the space of possibilities seems large.  Up to the present time, extragalactic SETI searches have focused mainly on detecting outliers to the Tully-Fisher relation (due to the presumed use of Dyson spheres)~\cite{annis1999b, zackrisson2015}, or in detecting extreme redness caused by waste heat~\cite{griffith2015}.  All have focused on nearby galaxies, examined individually rather than as a group.  Our geometric results should remain valid, independent of the specific galaxy-modification one might search for.

We organize this paper as follows:  Section 2 reviews the basic assumptions and equations that describe a cosmological aggressive expansion scenario.  In section 3, we obtain expressions for the shape and solid angle size of a single life-saturated domain as viewed from the Earth, and in section 4 we obtain expressions for the average distance at which a single visible appearance has taken place and its median angular size.  Section 5 derives an expression for the total fraction of the sky that should be eclipsed by at least one life-saturated domain. In section 6, we numerically evaluate these equations for a range of scenarios in which the expansion speed of such civilizations ranges between $.1c$ and $.9c$, and the appearance time of humanity ranges from the mean time of arrival to a two standard deviation latecomer.  Section 7 is a discussion of these results and their implications, including possible strategies for observation. 

\section{Review: Aggressive expansion of life in the universe}

In the context of a standard background cosmology\footnote{We use a spatially flat FRW solution with $ \Omega_{\Lambda 0} = .683$, $\Omega_{r 0} = 3 \times 10^{-5} $, $\Omega_{m 0} = 1-\Omega_{\Lambda 0} - \Omega_{r 0} $, and $H_0 = .069 \ Gyr^{-1} $, fixing the present age of the universe at $t_{0} \approx 13.75$ Gyr. We utilize co-moving coordinates, and units of Gyr and Gly for time and distance.  The scale factor is denoted by $a(t)$.}, an aggressive expansion scenario describes a subset of advanced life that wishes to maximize their access to physical resources in the universe.  We model the appearance of such life as spatially random events with a rate (per unit co-moving volume, per unit cosmic time) given by some function $f(t)$.  After appearing, each civilization expands spherically outward at some speed $v$ in the local co-moving frame of reference.  This reflects the use of spacecraft that spend a short time reproducing and seeding colonization at each galaxy (order one thousand or ten thousand years), followed by a relatively long time coasting between galaxies (order millions of years).  More exotic space travel technologies also give rise to the same constant-$v$ expansion model.  A more complete and general discussion of this model than we present here can be found in~\cite{olson2014, olson2015a}.  

In a general scenario, we can also model a time delay $T$ between the arrival of the ``probe front" of self-replicating spacecraft and the full saturation of galaxies, with any associated observable changes.  In extreme cases (when $v$ is very close to the speed of light and $T$ is very long), including $T$ leads to interesting effects on the observability of the process.  However, if the time delay is comparable to a typical galaxy colonization time that is driven by self-replicating spacecraft~\cite{valdes1980}, then $T$ will be unimportant for the cosmological process we describe, and it can be approximated as zero.  Similarly, a general description can include the presence of many types of expanders that utilize different $v$ and $T$, but it seems reasonable to assume that most aggressively expanding civilizations will arrive at similar practical limits to technology, leading to similar expansion velocities -- in the present discussion, we will thus restrict our attention to scenarios in which all expanders have the same expansion velocity.

When these simplifying assumptions are made, the unsaturated fraction of space in the universe, $g(t)$, is given by the Guth-Tye-Weinberg formula~\cite{guth1980, guth1981}:
\begin{eqnarray}
g(t) = e^{- \int_{0}^{t} f(t') V(t',t) dt' }
\end{eqnarray}
where the volume $V(t',t)$ of each expanding sphere is:
\begin{eqnarray}
V(t',t) =\frac{4 \pi}{3} \left( \int_{t'}^{t} \frac{v}{a(t'')} dt'' \right) ^3.
\end{eqnarray} 

A model for the appearance rate, $f(t)$, can be constructed by assuming that $f(t)$ should be proportional to the number of Earth-like planets that are between 4.5 Gyr and 6 Gyr old, that have not been exposed to a nearby gamma ray burst (GRB) in the last .2 Gyr.  The time dependence of $f(t)$ can thus be obtained to a reasonable approximation through astrophysical models, leaving most of the uncertainty to reside in the overall proportionality constant $\alpha$.  We express this model as
\begin{eqnarray}
f(t) = \alpha \, e^{-\int_{t-.2 Gyr}^{t} \frac{1.3}{Gyr} a(t')^{-2.1} \, dt'} \int_{t-6}^{t-4.5} PFR(t') \, dt'.
\end{eqnarray}
The exponential factor is the probability that a given Earthlike planet in the proper evolutionary time window has not experienced a GRB extinction event in the last .2 Gyr, assuming that GRB extinction events can be modeled as a Poisson process with intensity $\frac{1.3}{Gyr} a(t)^{-2.1}$.  The time dependence of these extinction events comes from the time dependence of the cosmic GRB rate found by Wanderman and Piran~\cite{wanderman2010}, while the current rate of GRB extinction events, $\frac{1.3}{Gyr}$, is chosen to reflect an analysis by Piran and Jimenez~\cite{piran2014} showing that the probability of a ``biospherically important event" (on Earth, or a planet in similar circumstances) due to GRB bursts is $\approx 50$\% in .5 Gyr.

The time dependence of the earthlike planet formation rate, $PFR(t)$, is obtained from a simplified version of a model due to Lineweaver~\cite{lineweaver2001}.  We set  $PFR(t) =N \, M(t) \, SFR(t)$, where $N$ is a normalization constant and $M(t)$ represents an average metallicity obtained by integrating the star formation rate $M(t) = \int_{0}^{t} SFR(t') \, dt'$, with $SFR(t)$ itself modeled by:
\begin{eqnarray}
   SFR(t) = \left\{
     \begin{array}{lcc}
       \frac{t}{3} 10^{t-3} & : &   t < 3  \\
       10^{-\frac{(t-3)}{10.75}} & : &  t \geq 3 .
     \end{array}
   \right.
\end{eqnarray} 
The normalization constant $N$ is chosen such that $PFR(t)$ has a maximum value of unity, i.e. our $PFR(t)$ is meant to describe only the cosmic time dependence of this appearance rate, with the relevant scale of the process being held by the proportionality constant $\alpha$ (which has units of appearances per $Gly^3$ per $Gyr$). 

Thus, for the purposes of this paper, we can specify an aggressive expansion scenario with two parameters, $v$ and $\alpha$.  While it is reasonable to guess at $v$ within a single order of magnitude (we analyze $.1 c \leq v \leq .9c$), there is vastly more uncertainty residing in $\alpha$, reflecting our lack of information regarding biological evolution in the universe and the behavior of advanced intelligence.

As a way to attach some physical intuition to our scenarios (which are clouded by the uncertainty inherent in $\alpha$) and their relative plausibility, we make use of the Self-Sampling Assumption, which asserts that, all else equal, we should reason as though we are randomly selected from the set of comparable observers who will ever have existed~\cite{bostrom2002}.  In our case, ``comparable observers" means early technological species still confined to their home planet \emph{who have not appeared within an already life-saturated galaxy}.  If we assume that the appearance rate for human-comparable life has the same cosmic time dependence in non-occupied space as the aggressive expanders, then the ``time of arrival distribution" for civilizations like ours will have the same time dependence as $g(t)  f(t)$.  

To reiterate the role of humanity and human-stage life in this analysis, we are not assuming that humanity is or will be an aggressively expanding civilization.  We also do not assume any particular fraction of human-stage life in the universe will become aggressive expanders.  We \emph{do} assume that the appearance rate for human-stage life has the same cosmic time dependence as $f(t)$, given by equation 2.3, but with a different proportionality constant (one larger than $\alpha$).  We then consider the set of all human-stage civilizations that will ever have existed.  The time of arrival distribution of this set will have the same time dependence as $g(t) f(t)$, i.e. the fraction of unsaturated space, times the appearance rate of expanders (up to some constant factor).  Implicit in this analysis is that both human-stage life and expanders appear with some quantifiable rate, i.e. that we are not unique unless the universe is finite and the appearance rate is sufficiently low that a single appearance of human-stage life (that being humanity) is plausible.

As shown in figure 1, if there were no aggressive expanders (i.e. $\alpha \rightarrow 0$, perhaps corresponding to a scenario in which the laws of physics do not allow for aggressive expansion), then our model has humanity appearing as a slight early-comer, relative to the mean time of arrival for other human-stage civilizations (the appearance rate for which does not need to go to zero along with the expanders).  This conclusion changes as $\alpha$ is increased, introducing expanders to the universe.  Depending on the values of $\alpha$ and $v$, humanity could have (for example) arrived at the mean ($\mu$) time of arrival for human-stage life, or as a 2 standard-deviation ($\sigma$) latecomer -- the Self-Sampling Assumption seems to imply that we need not focus our attention on scenarios where we are a more extreme latecomer, without some additional information to imply that ``all else is not equal" for the case of humanity.

\begin{figure}[!h]
\centering
\includegraphics[width=0.6\linewidth]{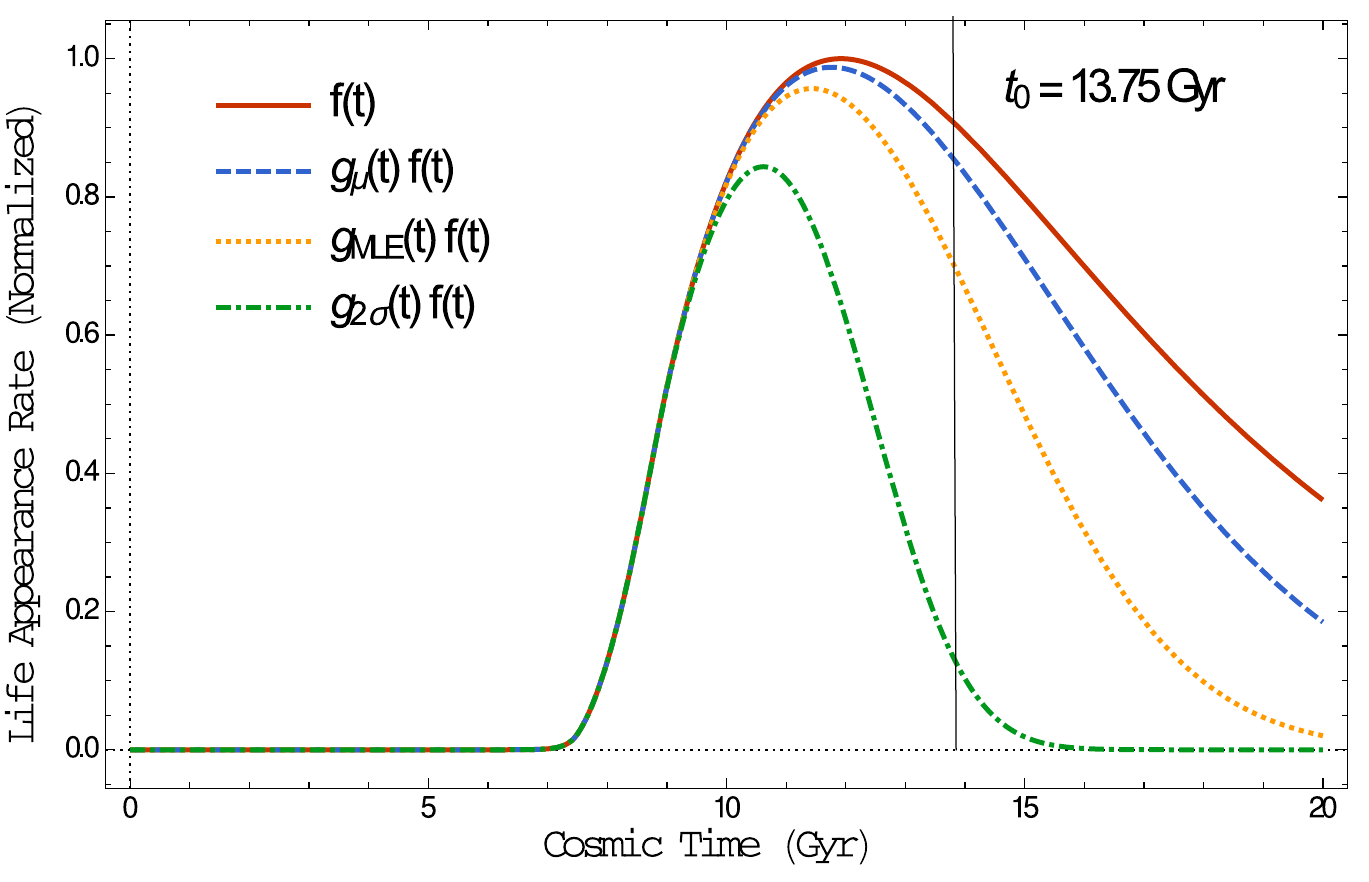}
\caption{Time dependence of $f(t)$ and $g(t) \, f(t)$ for three scenarios (with $v=.5c$). If no expanders are present, the appearance rate for human-comparable civilizations should be proportional to $f(t)$, and humanity has arrived slightly before the mean time of arrival for comparable life.  If expanders are present, then human-comparable life should appear at the rate $g(t) \, f(t)$.  The proportionality constant $\alpha$ that determines the appearance rate of expanders can be specified by the assumed time of arrival of humanity.  If humanity has arrived at the mean time of arrival or is a 2 standard deviation latecomer, then we label $g(t)$ as $g_{\mu}(t)$ or $g_{2 \sigma}(t)$, respectively.  For $\alpha$ chosen via the maximum likelihood estimate, we label $g_{MLE}(t)$.  }
\label{figure1}
\end{figure}

We can also find relevant values for $\alpha$ by making use of parameter estimation methods in the context of a single data point.  Here, we will make use of a maximum likelihood estimate (MLE) for $\alpha$ -- that is, the value of $\alpha$ that maximizes the probability that humanity should have appeared here at $t_0 = 13.75 Gyr$. I.e. we find the value of $\alpha$ to maximize $g(t_0) f(t_0) / \int_{0}^{\infty} g(t) f(t) \, dt$.

Because we have more intuition for the amount of luck involved in our appearance time than for the magnitude of $\alpha$ itself, we will specify scenarios by giving $v$ and the relative time of our arrival (or use of MLE), using a numerical search to find the corresponding value of $\alpha$.   

\section{Visible shape and size of a single life-saturated domain}

Suppose an aggressively expanding civilization appears at a co-moving coordinate distance $R$ from the Earth, at cosmic time $t_{app}$.  It expands in a sphere about its point of origin, but the expanding boundary will not appear as a sphere from the Earth at time $t_0$, due to the finite speed of light.  This type of geometry has been studied since the 1960's in the context of relativistic explosions~\cite{rees1967,ginzburg1979}, but in our case we cannot generally rely on the approximation that the distance $R$ is much, much larger than the size of the domain itself, and thus we will give exact expressions here and use them in the following sections.

\begin{figure}[!h]
\centering
\vspace{3mm}
\includegraphics[width=0.6\linewidth]{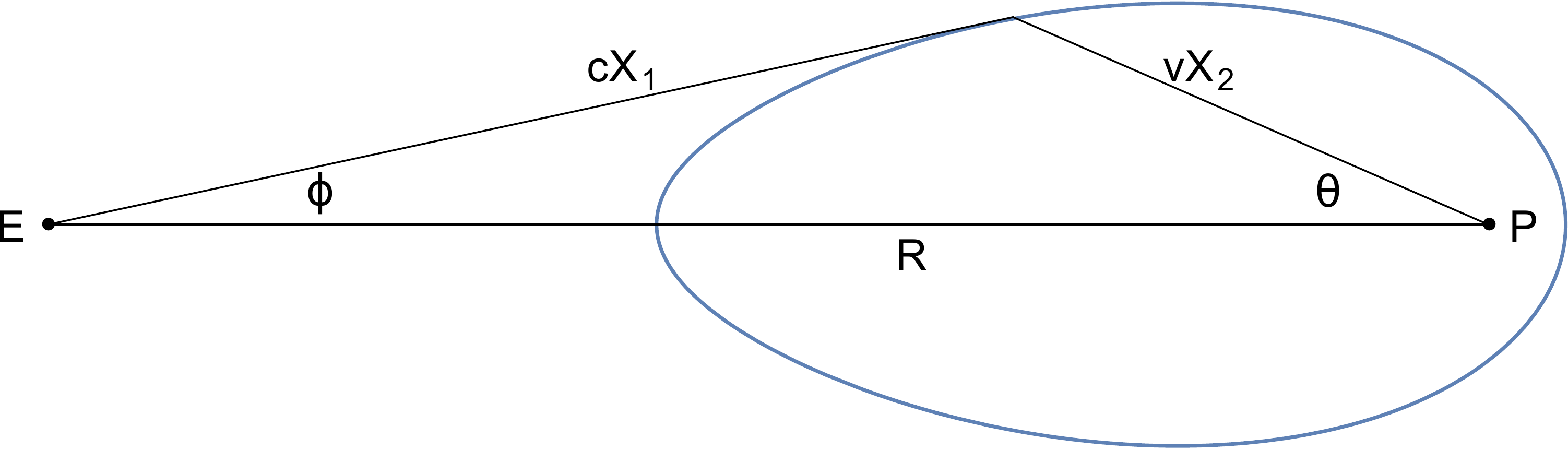}
\caption{Visible geometry of a civilization expanding from point $P$ a coordinate distance $R$ away from the observation point point $E$ (the Earth).  The visible domain boundary in this case corresponds to an expansion velocity of $v = .85 c$. }
\label{figure2}
\vspace{0mm}
\end{figure}

The geometry of the situation is depicted in figure 2.  The coordinate distance from the Earth to the point of observed light emission at the domain boundary is $c X_1$, where $X_1 = \int^{t_{0}}_{t_{e}} \frac{1}{a(t')} \, dt'$.  The coordinate distance from the origin of the expanding civilization to the observed point of emission is $v X_2$, where $X_2 = \int^{t_{e}}_{t_{app}} \frac{1}{a(t')} \, dt'$.  We also introduce the quantity $X = X_1 + X_2 =  \int^{t_{0}}_{t_{app}} \frac{1}{a(t')} \, dt'$, which eliminates dependence on the emission time $t_e$.  In the interest of compact notation, we will suppress $X$'s dependence on $t_0$ (which we assume to be a constant $13.75 Gyr$), writing either $X$ or $X(t)$, with $t$ referring to the domain's appearance time $t_{app}$.

Using the law of cosines and the definition of $X$ allows us to solve for the distance $v X_2$ as a function of $R$, $X$, and the expansion angle $\theta$:
%\begin{widetext}

\begin{eqnarray}
v X_2 = \frac{v \left( R v \cos (\theta ) -c^2 X + \sqrt{R v \cos (\theta ) \left(R v \cos (\theta )-2 c^2 X\right)+c^2 v^2 X^2+R^2 (c^2 - v^2)}\right)}{v^2-c^2}.
\end{eqnarray}

This encodes the visible shape of these domains, as viewed from the Earth.  Using the law of sines to eliminate $X_2$ allows us to express $\phi$ as a function of $R$, $X$, and $\theta$:

\begin{eqnarray}
\resizebox{.92 \textwidth}{!}{$
\phi = \tan ^{-1}\left(\frac{2 v (c^2 X^2 - R^2)}{R \csc (\theta ) \left(2 c^2 X + \sqrt{4 c^2 \left(R^2+v^2 X^2\right)+2 R v \left(R v \cos (2 \theta )-4 c^2 X \cos (\theta )\right)-2 R^2 v^2}\right)-2 v c^2  X^2 \cot (\theta )}\right).$}
\end{eqnarray}

We will be interested in the visible size of such a domain, and thus we are interested in the maximum of $\phi$ over $\theta$, which we label as $\phi_{max}$ and express as:

\begin{eqnarray}
\phi_{max}= \tan ^{-1}\left(\frac{v (c^2 X^2 - R^2) \sqrt{(R^2 - v^2 X^2)}}{c^2 X (R^2 - v^2 X^2)+R^2 \sqrt{(c^2-v^2) (R^2-v^2 X^2)}}\right).
\end{eqnarray}

As viewed from the Earth, the solid angular size $\Omega$ of this domain, as a fraction of the total celestial sphere ($4 \pi$), is therefore $\frac{1}{2}(1 - \cos(\phi_{max}))$, i.e.

\begin{eqnarray}
\frac{\Omega}{4 \pi }= \frac{c R-\sqrt{c^2 (R^2-v^2 X^2)+v^2 \left(2 X \left(\sqrt{(c^2-v^2) (R^2-v^2 X^2)}+v^2 X\right)-R^2\right)}}{2 c R}.
\end{eqnarray}

%\end{widetext}

\section{Median angular size and distance to life-saturated domains}

A civilization appearing a distance $R$ away (say, on the order of a Gly) must have appeared within a specific time window if it is visible to us.  The earliest possible time is set by the condition $R = v X$ -- appearing any earlier means that this civilization would have overtaken us by now and saturated our galaxy with life.  The latest possible time is set by $R = c X$ -- appearing any later means that the light from this civilization has not had time to reach us.  

The present visible size $\Omega$ of this civilization is at its maximum\footnote{Note that $\Omega$ only reaches a maximum of half the celestial sphere in the limit that $v \ll c$.  For expansion speeds that are a substantial fraction of $c$, the maximum value of $\Omega$ will be smaller.} if the civilization appeared at the beginning of the window, and will be vanishingly small if it appeared at the end of the window.  As a function of the civilization's appearance time, $\Omega$ decreases monotonically between the opening of the window and its closing.  The appearance probability, however, is governed by $f(t)$, which can change significantly over the course of Gyr.  In cases where the window is open for several Gyr (imagine a civilization $1 Gly$ away, expanding at $.1 c$), it can be dramatically more likely that a civilization will appear towards the end of the window (thus appearing small) than appear at the beginning.

We wish to compute the median solid angle size of a civilization, given that the civilization appeared at co-moving distance $R$.  Because of the monotonic decrease in angular size with appearance time, this is equivalent to computing the solid angle size of a civilization appearing at the median appearance time within the visibility time window.  The median time for a visible appearance at $R$ is found by numerically solving the following for $t_{med}$:
%\begin{widetext}
\begin{eqnarray}
\resizebox{.92 \textwidth}{!}{$
\int_{0}^{t_{med}} \Theta(R - v X(t_{app})) \, \Theta(c X(t_{app})-R) \, f(t_{app}) \, dt_{app} = \frac{1}{2} \int_{0}^{t_{0}} \Theta(R - v X(t_{app})) \, \Theta(c X(t_{app})-R) \, f(t_{app}) \, dt_{app}$}
\end{eqnarray} 
%\end{widetext}
where $\Theta$ is the Heaviside step function, which is used to ensure that the integral is only over the observable time window.  Having calculated $t_{med}$ for a given $R$ within a given scenario, one can now compute $\Omega_{med}$ by substituting $R$ and $X = \int^{t_{0}}_{t_{med}} \frac{1}{a(t)} \, dt$ into equation 3.4.  

$\Omega_{med}$ gives us the median solid angle size of a visible civilization, \emph{given that it appeared at co-moving distance $R$}.  However, it may be that an appearance at $R$ is very improbable -- for any given scenario, there will always be a range of $R$ for which this is the case.  Thus, if we determine and plot $\Omega_{med}(R)$ for a given scenario, we should also specify the distance $R_1$ at which one visible expanding civilization, on average, will appear\footnote{$R_1$ is defined differently here, compared to ref~\cite{olson2015a}.  There, $R_1$ was the distance to the \emph{leading edge} of the nearest expanding civilization (on average), whereas it here represents the distance to the nearest appearance location (on average).}.   This will give us an idea of the characteristic angular size of the nearest visible civilizations, for a given scenario. 

To calculate $R_1$, we use the following approximation for the average number of observable civilizations appearing within co-moving distance $R$:
\begin{eqnarray}
EV_{R}(obs) = \int_{0}^{t_0} f(t) \, V_{R}(t,t_0)  \, dt
\end{eqnarray}
where $V_{R}(t,t_0)$ is the co-moving volume of space within our past light cone at time $t$ (to a maximum distance of $R$), minus the co-moving volume of space within our ``past saturation cone" (to a maximum distance of $R$).  I.e. we can express $ V_{R}(t,t_0)$ as:
%\begin{widetext}
\begin{eqnarray}
 V_{R}(t,t_0)= \frac{4 \pi}{3}\left[ (c X(t))^3 \, \Theta(R - c X(t)) + R^3 \, \Theta(c X(t) - R ) \right] \nonumber \\ 
 - \frac{4 \pi}{3}\left[(v X(t))^3 \,  \Theta(R - v X(t)) + R^3 \, \Theta(v X(t) - R )\right].
\end{eqnarray}
%\end{widetext}

Obtaining $R_1$ for a given scenario amounts to numerically finding the solution to $EV_{R_1}(obs)=1$. 

This form of $EV_{R}(obs)$ is an approximation, because it counts ``virtual bubbles" (expanding civilizations appearing within already expanding civilizations) as independently observable events.  For the purposes of calculating $R_1$, it should be an excellent approximation.  Since this formula implicitly assumes that appearances are independent of one another in space and time (a Poisson process), $R_1$ may also be interpreted as the distance at which the probability of having \emph{zero} visible appearances is $e^{-1} \approx 37 \%$.

\section{Fraction of celestial sphere eclipsed by at least one life-saturated domain}

The previous section shows how we can calculate the characteristic observable size of life-saturated domains in an aggressive expansion scenario.  We can also calculate the average total fraction of the celestial sphere eclipsed by at least one domain.  To do this, we first find the co-moving coordinate volume of space (at a particular cosmic time, $t$) that is able to produce a domain that will eclipse the celestial sphere along a particular line of sight from an observer at time $t_0$, denoted $V_{LOS}(t)$.  From the previous section, we know that visible domains can appear at cosmic time $t$ along the line of sight extending from $R = v X(t)$ to $R = c X(t)$.  If a domain appears within angle $\phi_{max}$ of the line of sight, it will also be eclipsed.  Thus, we can express $V_{LOS}(t)$ as:

\begin{eqnarray}
V_{LOS}(t) &=& 2 \pi \int_{v X(t)}^{c X(t)} \int_{0}^{\phi_{max}(R,t)} R^2 \, \sin(\phi) \, d \phi \, d R  \\
&=& 2 \pi  \int_{v X(t)}^{c X(t)} R^2 \left[ 1-\cos(\phi_{max}(R,t)) \right] d R \\
&=& \frac{\pi v^2}{3 c} (c-v)^2 X(t)^3
\end{eqnarray}
where equation 3.3 was used for the final integration. 

Now express the probability that \emph{zero} expanding civilizations have appeared between time $t_i$ and $t_f$ and eclipse the line of sight as $k(t_i,t_f)$ -- this will satisfy $k(t_i,t_f) = k(t_i,t_m) \, k(t_m,t_f)$ for a middle time $t_m$.  This allows us to write:
\begin{eqnarray}
k(0,t+dt) = k(0,t) \, k(t,t + dt)\\
 = k(0,t) \, [1 - f(t) \, V_{LOS}(t) \, dt]\\
 \frac{d k(0,t)}{dt} = -k(0,t) \, f(t) \, V_{LOS}(t).
\end{eqnarray} 
Integrating gives:
\begin{eqnarray}
k \equiv k(0,t_0) = e^{- \int_{0}^{t_0} f(t) \, V_{LOS}(t) \, dt }
\end{eqnarray}
which represents the total probability that \emph{no} domains eclipse the celestial sphere along the line of sight.  The probability that \emph{at least one} domain eclipses the line of sight is thus $1 - k$.  Due to the compact but exact form of $V_{LOS}(t)$, we can also express $k$ as:
\begin{eqnarray}
k = e^{- \frac{\pi v^2 (c-v)^2}{3c} \int_{0}^{t_0} f(t) \, X(t)^3 \, dt }.
\end{eqnarray}

\section{Numerical results:  Angular sizes, distances, and the eclipsed fraction of the sky}

While the median visible angular size of an expanding civilization appearing at $R$ depends on $f(t)$, only the time dependence of $f(t)$ is important -- the overall proportionality constant $\alpha$ does not effect $\Omega_{med}(R)$, as one can verify from equation 4.1.  However, the distance to the nearest visible expanding domain will certainly depend on $\alpha$.  We will quantify this dependence by calculating $R_1$ for any given scenario, i.e. the co-moving distance at which the expected value of the number of visible expanding civilizations is equal to unity, according to equations 4.2 and 4.3.

In each component of figure 3 appearing below, we plot $\Omega_{med}(R) / 4 \pi$ (i.e. the median fraction of the total celestial sphere eclipsed by a civilization appearing at $R$), and simultaneously give up to three values for $R_1$, corresponding to humanity appearing at the mean time of arrival $R_1 (\mu)$, humanity appearing as a two standard deviation latecomer $R_1(2 \sigma)$, or the maximum likelihood estimate $R_1(MLE)$.  For some expansion scenarios, some of these $R_1$ will not be defined, and thus do not appear in the plot.  $R_1$ will not be defined when the expected value of the number of visible domains is less than 1 for all values of $R$ -- this tends to happen in scenarios with high expansion velocity. 

Figure 4 is a combined plot of $\Omega_{med}(R) / 4 \pi$ for each expansion velocity, for a more direct comparison without reference to $R_1$.

In table 1, we give the quantity $1-k$, i.e. the average fraction of the celestial sphere eclipsed by at least one domain, for 27 scenarios, corresponding to expansion speeds from $.1 c$ to $.9 c$ and the same three relative times of arrival, $\mu$, $\mu + 2 \sigma$, and MLE.  Notice the distinction between $\Omega_{med}(R_1) / 4 \pi$ and the quantity $1-k$ -- the former represents the characteristic fraction of the sky eclipsed by the closest visible domain, while the latter represents the cumulative fraction of the sky eclipsed by all domains in a given scenario.

\begin{figure}[!h]
\centering
\subfloat[]{
  \includegraphics[width=0.45\linewidth]{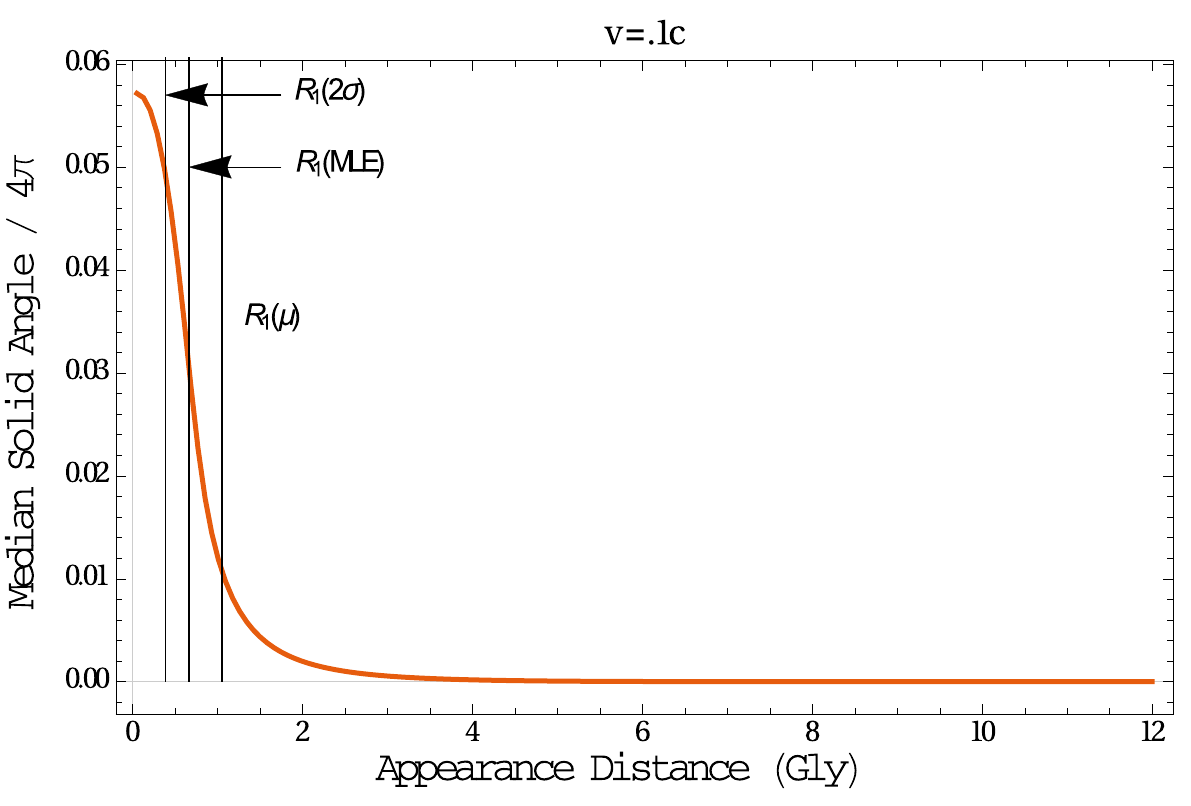}
}
\subfloat[]{
  \includegraphics[width=0.45\linewidth]{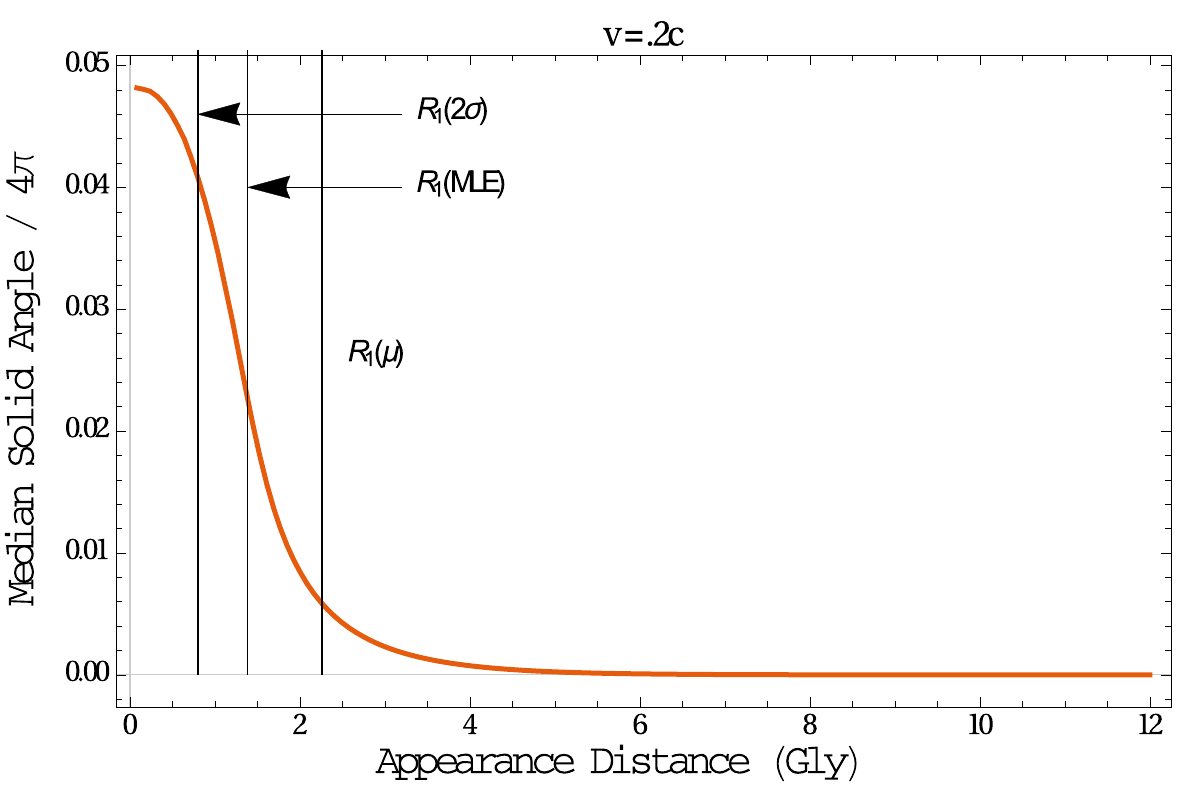}
}
\hspace{0mm}
\subfloat[]{
  \includegraphics[width=0.45\linewidth]{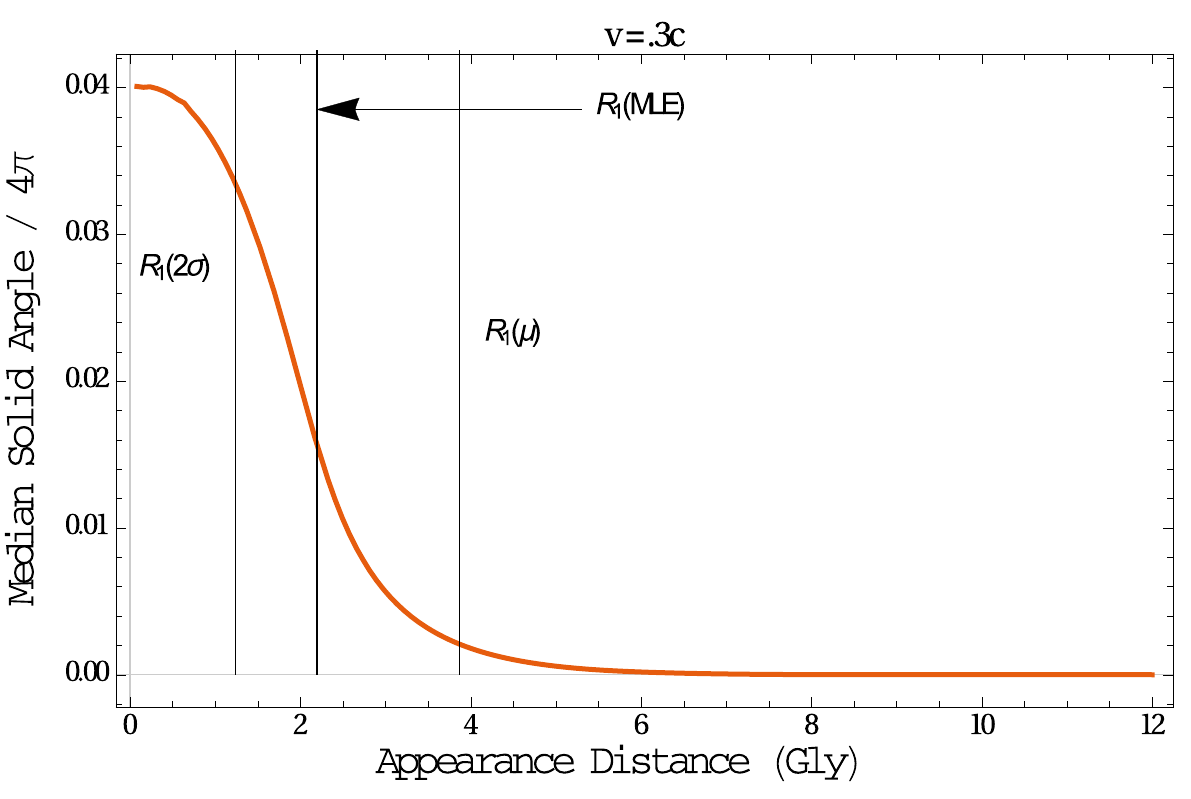}
}
\subfloat[]{
  \includegraphics[width=0.45\linewidth]{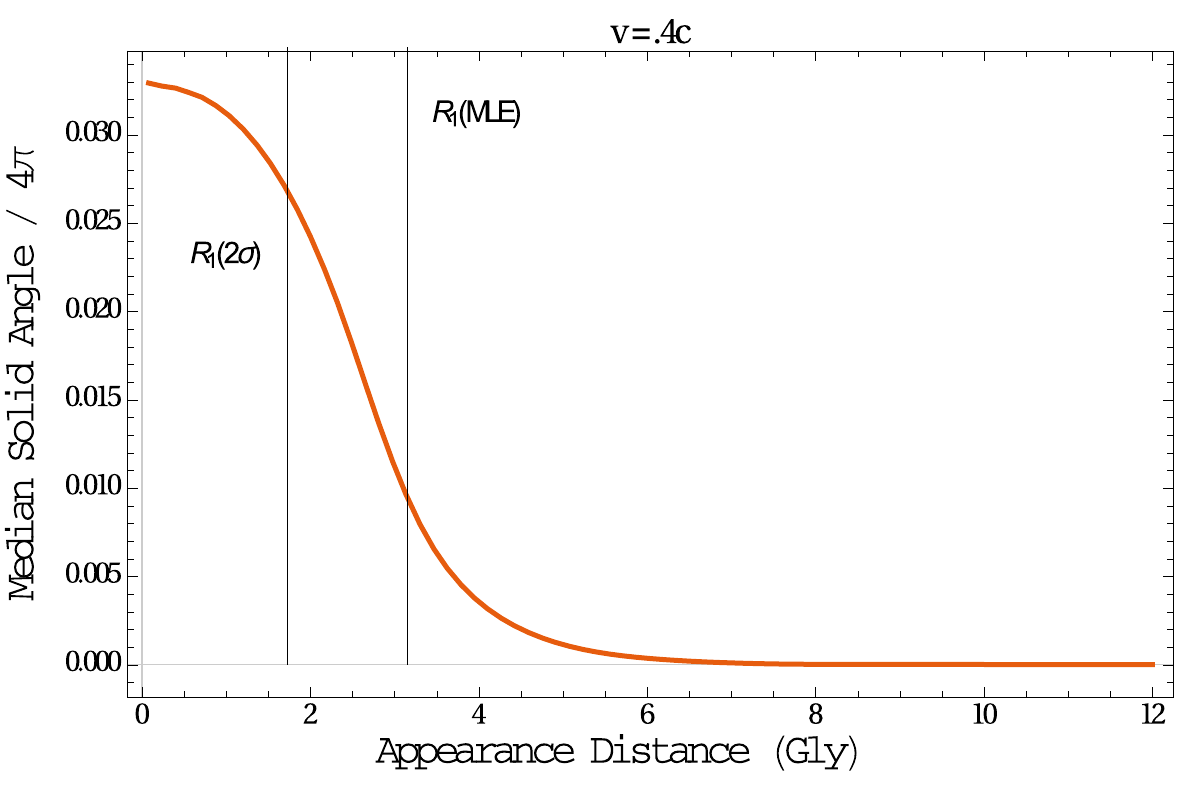}
}
\hspace{0mm}
\subfloat[]{
  \includegraphics[width=0.45\linewidth]{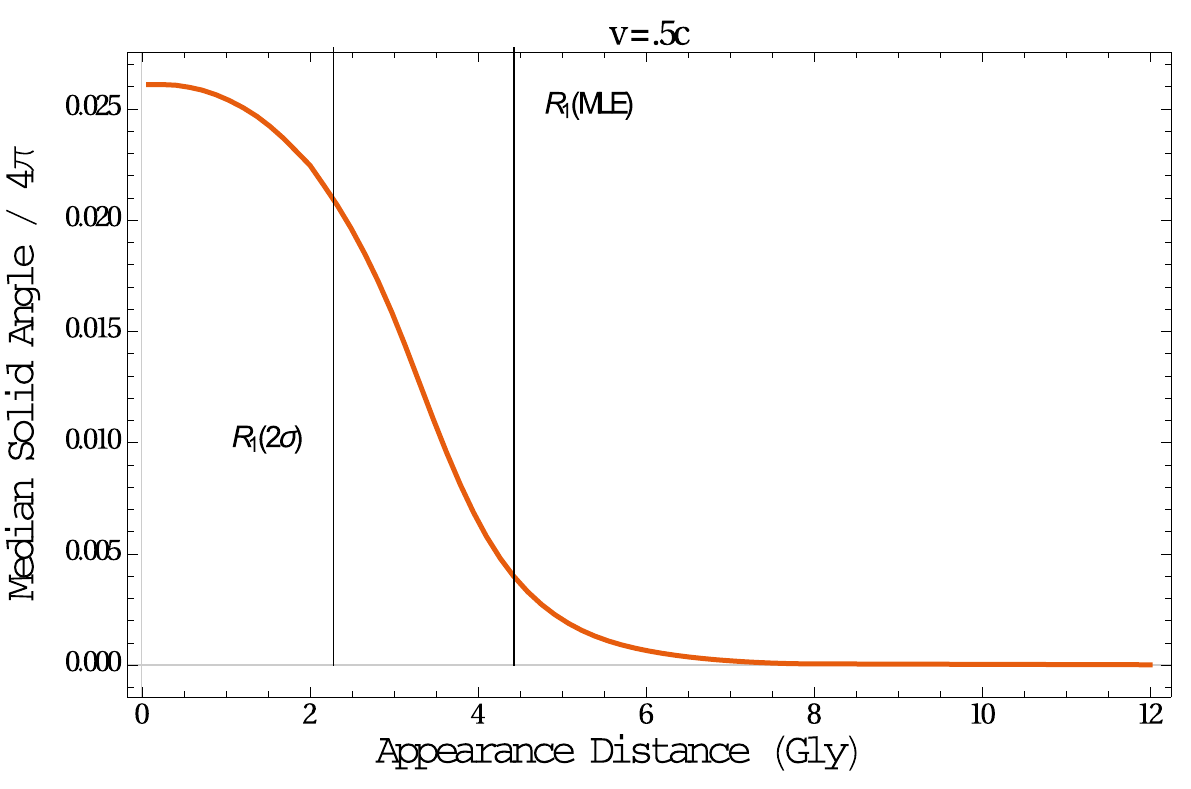}
}
\subfloat[]{
  \includegraphics[width=0.45\linewidth]{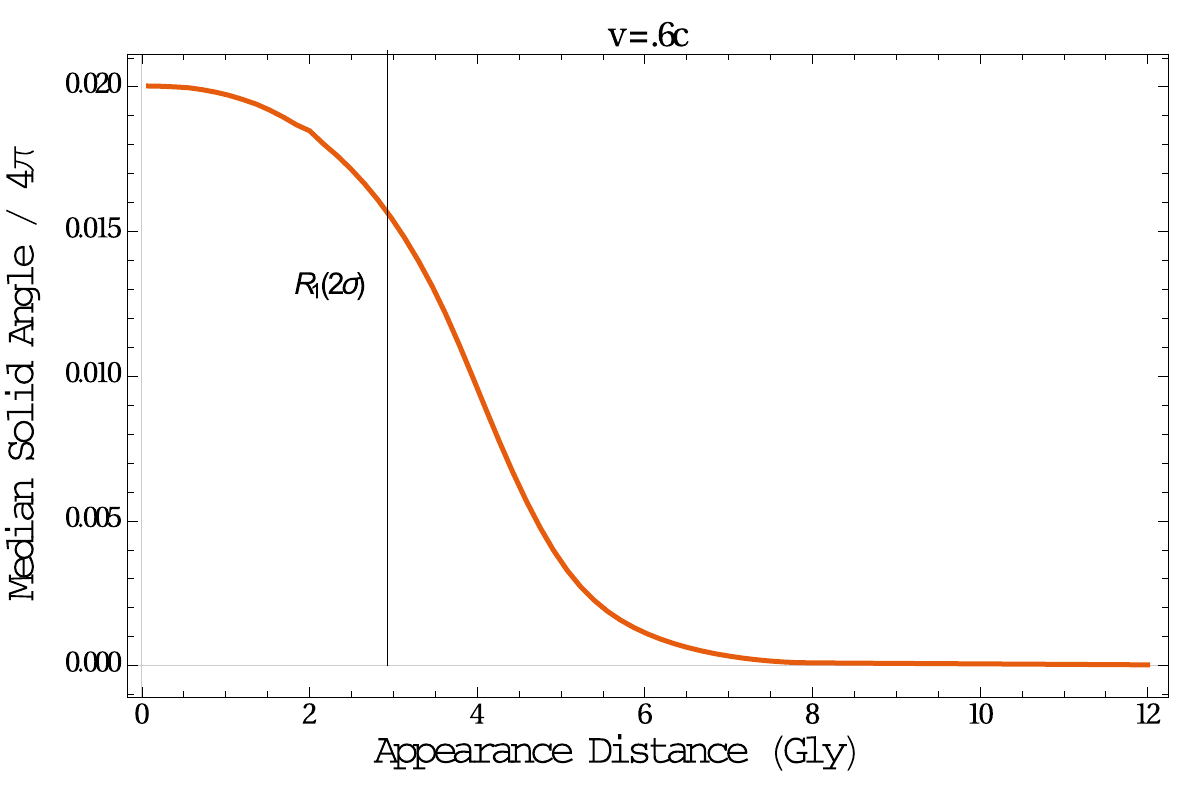}
}
\caption{}
\end{figure}

\cleardoublepage

\begin{figure}
\centering
\ContinuedFloat
\subfloat[]{
  \includegraphics[width=0.45\linewidth]{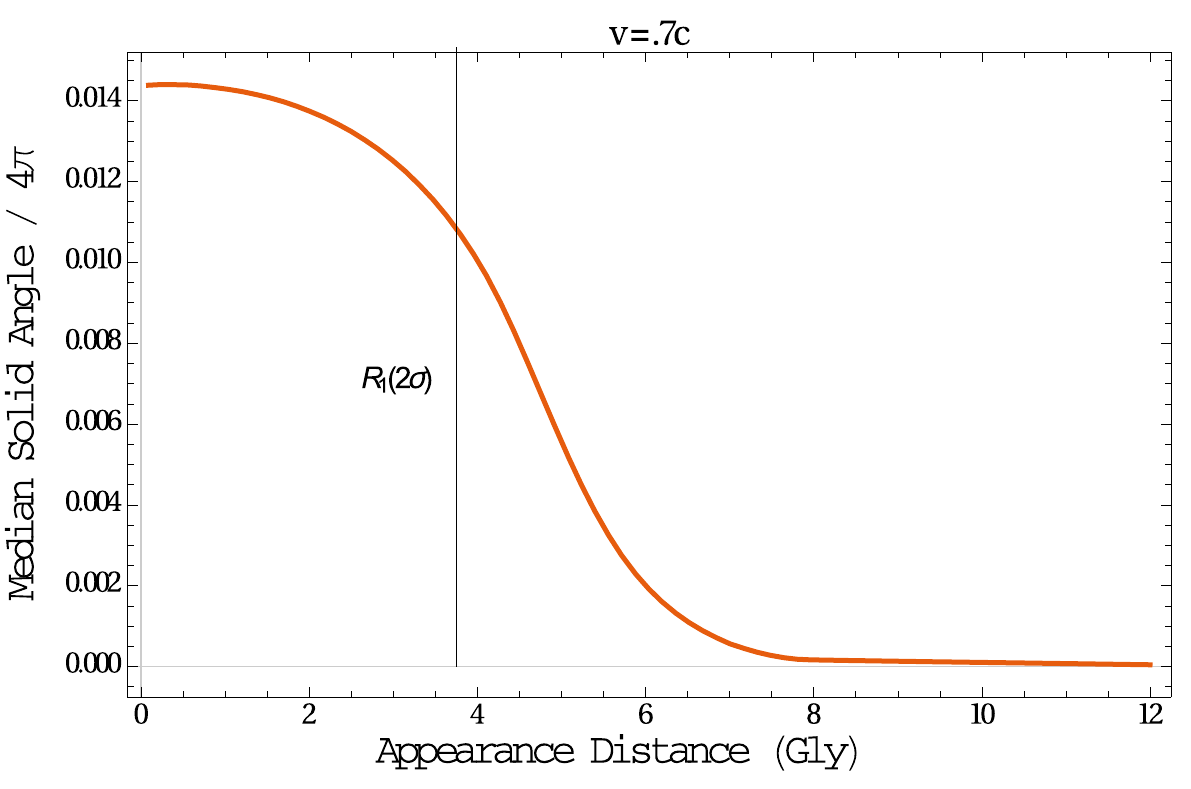}
}
\subfloat[]{
  \includegraphics[width=0.45\linewidth]{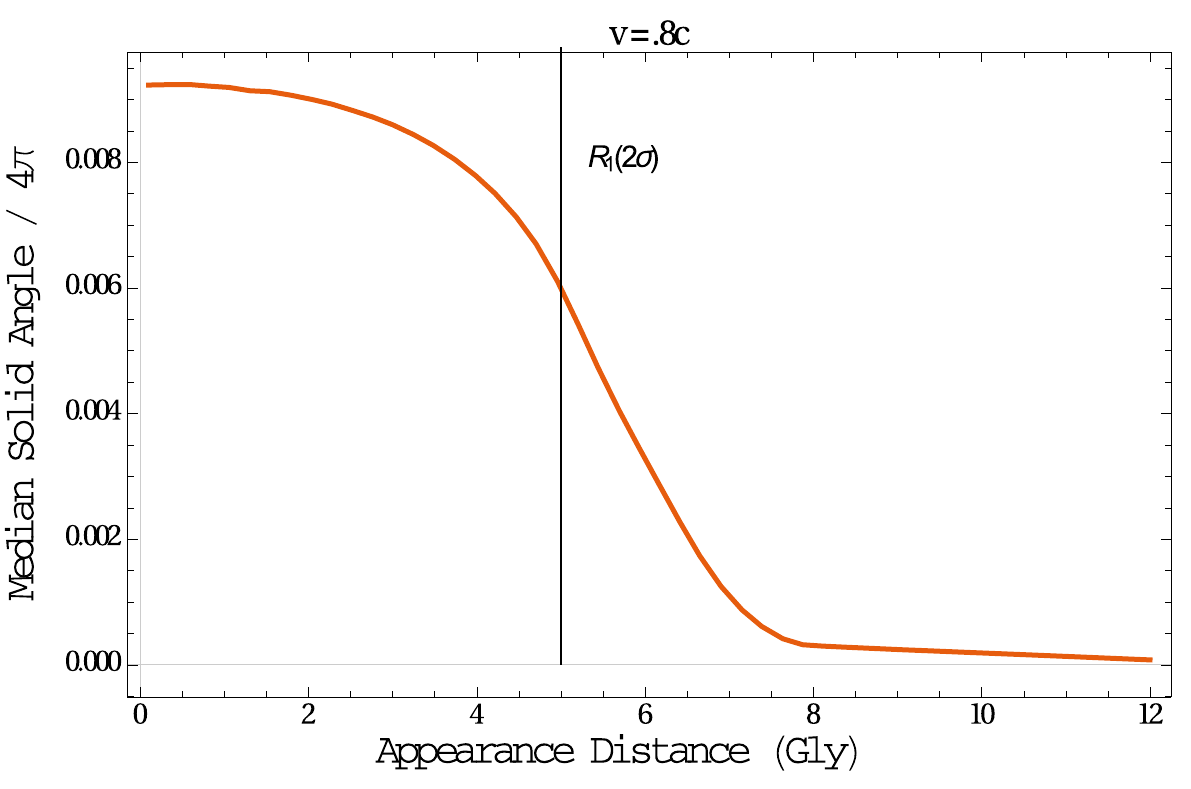}
}
\hspace{0mm}
\subfloat[]{
  \includegraphics[width=0.45\linewidth]{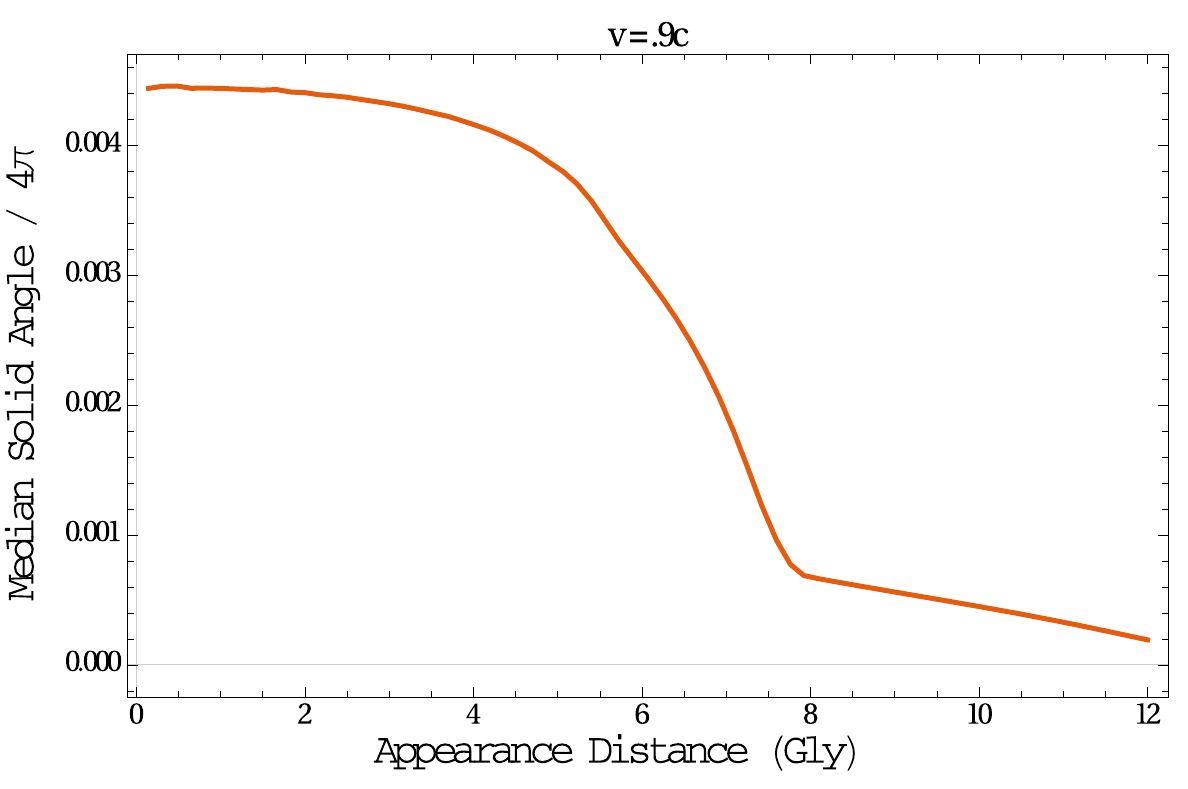}
}
\caption{Numerically generated plots of $\Omega_{med} / 4 \pi$ vs. $R$ for nine expansion velocities.  In each plot, $R_1$ (the distance to one appearance, on average) is marked with a vertical line for up to three scenarios, in which humanity has arrived at the mean time of arrival $R_1(\mu)$, as a two standard deviation latecomer $R_1(2 \sigma)$, or is the prediction of the Maximum Likelihood Estimate $R_1(MLE)$.}
\end{figure}

\begin{figure}[!h]
\centering
\vspace{3mm}
\includegraphics[width=0.6\linewidth]{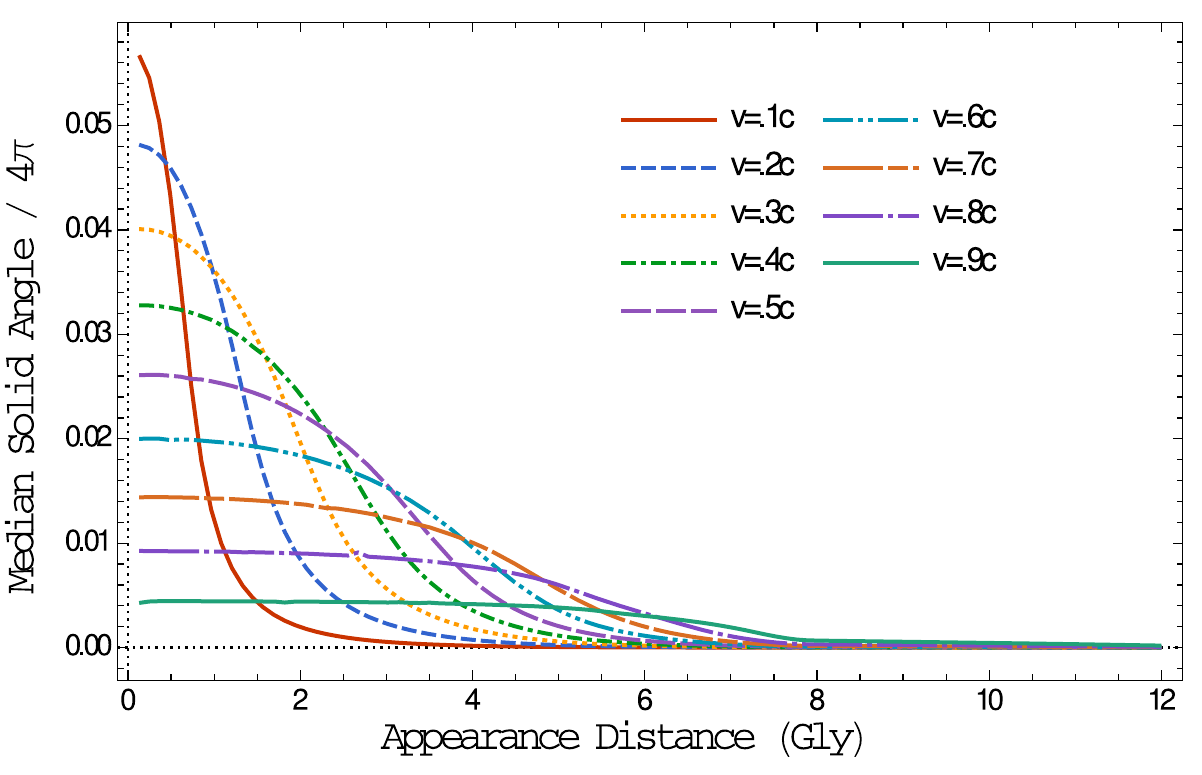}
\caption{All $\Omega_{med} / 4 \pi$ vs. $R$ for direct comparison. Note that higher expansion velocity scenarios produce smaller visible domains nearby, but larger visible domains at long distance.}
\label{figure3}
\vspace{0mm}
\end{figure}

\begin{table}[!h]
\centering
\resizebox{\linewidth}{!}{$
\begin{tabular}{|c||c|c|c|c|c|c|c|c|c|}
\hline Expansion Speed  & $v=.1c$ & $v=.2c$  & $v=.3c$ & $v=.4c$  & $v=.5c$  & $v=.6c$ & $v=.7c$ & $v=.8c$ & $v=.9c$ \\ 
\hline $1-k$ (humanity at $\mu$) & .11 & .045 & .023 & .013 & .0072 & .0038 & .0018 & .00072 & .00016 \\ 
\hline $1-k$ (Max. Likelihood Est.) & .39 & .18 & .096 & .054 & .030 & .016 & .0079 & .0031 & .00069 \\ 	
\hline $1-k$ (humanity at $\mu + 2 \sigma$) & .98 & .77 & .53 & .34 & .21 & .12 & .058 & .023 & .0051 \\ 
\hline 
\end{tabular} $}
\caption{Average fraction of the sky covered by at least one life-saturated domain ($1-k$) for 27 scenarios, depending on the expansion speed of the domains, and the appearance rate constant $\alpha$, set by the assumed relative time of arrival of humanity or the Maximum Likelihood Estimate.}
\end{table}

%\cleardoublepage

\section{Discussion of results and conclusions}
From the previous plots, our main result is immediate -- if we can observe expanding, life-saturated domains, then they \emph{will likely appear at great distance but be very large in the sky.}  For example, in the maximum likelihood estimate scenarios, the characteristic angular size $\Omega_{med}(R_1) / 4 \pi$ can be seen to range from $2.7 \%$ of the sky in a $v=.1c$ scenario to $.22 \%$ of the sky if the expansion velocity is $v=.5c$.  As we have pointed out before~\cite{olson2015a}, $R_1$ is a cosmological distance (larger than the homogeneity scale of the universe) in nearly all cases, and the fastest expansion speed scenarios are unlikely to have produced even a single visible life-saturated domain, from our present vantage point.

We also note that if one expects to see aggressive expanders at all (i.e. when $R_1$ is defined), then the average fraction of the sky covered by \emph{at least one domain} ranges from a few percent, at minimum, to nearly completely filled.  Although some of these domains can be expected to be very distant, the life appearance rate does not become appreciable until a cosmic time of $\approx 7.5$ Gyr, corresponding to a maximum visible appearance distance of $\approx 8$ Gly or a redshift of $\approx .66$, and thus virtually all domains should appear closer than this.

These results have implications for search strategies.  It would be technically challenging and costly to conduct a full-sky survey capable of resolving and analyzing every galaxy in detail out to a sufficient distance, due to the magnitude of the distances involved.  Such an approach would correspond to a scaled-up version of the largest Kardashev type III galaxy search so far~\cite{griffith2015} (an analysis of galaxies in the WISE survey data).  However, the large median size of the domains suggest search strategies that offset this difficulty.  One could imagine, for example, a series of very narrow but sufficiently deep surveys, distributed over the sky, that could analyze galaxies for signs of life at the necessary distance.  In this type of search, one would hope to penetrate a domain and capture a small number of modified galaxies as a first observation, rather than capture a life-saturated domain in its entirety.

We have focused here on the geometry of expanding domains, but we have not discussed the observable signatures of the individual life-saturated galaxies that compose them (see~\cite{wright2014b} for a treatment of this issue).  However, we should point out one correlation that seems likely:  Generally higher limits to practical technology should allow for a greater degree of galaxy-engineering and energy use, and also for greater expansion velocity.  Since faster expansion scenarios produce fewer observable domains from our current vantage point, we expect a trade-off in observability.  Lower-technology scenarios can produce more and closer domains, expanding slowly, that are more difficult to discern as life-saturated.  Higher-technology scenarios produce fewer and more distant domains that expand more rapidly, but their galaxies should more clearly stand out as engineered.  The highest technology scenarios of all (some of which could eventually perturb the evolution of the universe by releasing enormous quantities of waste heat~\cite{olson2014}) are exactly those for which we expect such rapid expansion that $R_1$ is not defined.  In other words, the most radical scenarios for the universe are those most likely to leave us, for the time being, with a universal Fermi paradox.

\bibliographystyle{JHEP}
\bibliography{ref5}

\end{document}